\documentclass[prl,superscriptaddress,twocolumn,showpacs]{revtex4}
\usepackage{epsfig}
\usepackage{graphicx}
\usepackage{amsmath}
\usepackage{amssymb}
\usepackage{color}

\begin{document}

\title{Dirac semimetals $\mathbf{A_3 Bi}$ ($\mathbf{A}=\mathbf{Na},\mathbf{K},\mathbf{Rb}$) as $\mathbb{Z}_\mathbf{2}$ Weyl semimetals}
\date{\today}

\author{E. V. Gorbar}
\affiliation{Department of Physics, Taras Shevchenko National Kiev University, Kiev, 03680, Ukraine}
\affiliation{Bogolyubov Institute for Theoretical Physics, Kiev, 03680, Ukraine}

\author{V. A. Miransky}
\affiliation{Department of Applied Mathematics, Western University, London, Ontario N6A 5B7, Canada}

\author{I. A. Shovkovy}
\affiliation{College of Letters and Sciences, Arizona State University, Mesa, Arizona 85212, USA}

\author{P. O. Sukhachov}
\affiliation{Department of Physics, Taras Shevchenko National Kiev University, Kiev, 03680, Ukraine}

\begin{abstract}

We demonstrate that the physical reason for the nontrivial topological properties of Dirac
semimetals $\mathrm{A_3Bi}$ ($\mathrm{A}=\mathrm{Na},\mathrm{K},\mathrm{Rb}$)
is connected with a discrete symmetry of the low-energy effective Hamiltonian.
By making use of this discrete symmetry, we argue that all electron states can be split
into two separate sectors of the theory. Each sector describes a Weyl semimetal with
a pair of Weyl nodes and broken time-reversal symmetry. The latter symmetry
is not broken in the complete theory because the time-reversal transformation interchanges
states from different sectors. Our findings are supported by explicit calculations of the Berry
curvature. In each sector, the field lines of the curvature reveal a pair of monopoles of the
Berry flux at the positions of Weyl nodes. The $\mathbb{Z}_2$ Weyl semimetal nature is
also confirmed by the existence of pairs of surface Fermi arcs, which originate from different
sectors of the theory.
\end{abstract}



\maketitle

{\em Introduction.} Three-dimensional (3D) Dirac semimetals whose conduction
and valence bands touch only at discrete (Dirac) points in the Brillouin zone with
the electron states described by the 3D massless Dirac equation are 3D analogs
of graphene. Historically, bismuth \cite{Cohen} was the first material where it was
shown that its low-energy quasiparticle excitations near the $L$ point of the
Brillouin zone are described by the 3D Dirac equation with a small mass \cite{Wolff}.
Since the Dirac point is composed of two Weyl nodes of opposite chirality which
overlap in momentum space, it can be gapped out. Therefore, even if the 3D Dirac
point is obtained accidentally by fine tuning the spin-orbit coupling strength or
chemical composition, it is, in general, not stable and is difficult to control.

It was proposed in Refs.~\cite{Manes,Mele} that an appropriate crystal symmetry
can protect and stabilize the 3D Dirac points if two bands which cross each other
belong to different irreducible representations of the discrete crystal rotational
symmetry. By using the first-principles calculations and effective model analysis,
$\mathrm{A_3Bi}$ (A=Na, K, Rb) and $\mathrm{Cd_3As_2}$ compounds were
identified in Refs.~\cite{Fang,WangWeng} as 3D Dirac semimetals protected by
crystal symmetry. Various topologically distinct phases can be realized in these
compounds by breaking time-reversal and inversion symmetries. By making use
of angle-resolved photoemission spectroscopy, a Dirac semimetal band structure
was indeed observed \cite{Borisenko,Neupane,Liu} in $\mathrm{Cd_3As_2}$ and
$\mathrm{Na_3Bi}$, opening the path toward the experimental investigation of the 
properties of 3D Dirac semimetals. For a recent review of 3D Dirac semimetals, see
Ref.~\cite{Gibson:1411.0005}.

Closely related to 3D Dirac semimetals are Weyl semimetals. They were proposed
to be realized in pyrochlore iridates \cite{Savrasov}, topological heterostructures
\cite{Balents}, and magnetically doped topological insulators \cite{Cho}. Although
not experimentally observed yet, Weyl semimetals have been very actively studied
theoretically (for reviews, see Refs.~\cite{Hook,Turner,Vafek}). A Weyl node is
topologically nontrivial because it is a monopole of the Berry flux in momentum
space. This is also the reason why Weyl nodes can appear or annihilate only in
pairs \cite{Nielsen}.

The simplest way to turn a Dirac semimetal into a Weyl one is to apply an external
magnetic field, which breaks time-reversal symmetry. This can be realized
even in high-energy physics context \cite{GMS2009}. In fact, the corresponding
transition might have already been observed in $\mathrm{Bi_{1-x}Sb_x}$ for
$x \approx 0.03$ \cite{Kitaura} and in $\mathrm{Cd_3As_2}$ \cite{1412.0824}.
In the case of $\mathrm{Bi_{1-x}Sb_x}$ \cite{Kitaura}, the authors measured
negative magnetoresistivity at not very large magnetic fields that might be a
fingerprint of a Weyl semimetal phase \cite{Nielsen,Son} (see, however, the
discussion in Ref.~\cite{magnetoresistivity}). In the case of $\mathrm{Cd_3As_2}$,
a magnetic field driven splitting of Landau levels consistent with the Weyl phase,
time-reversal symmetry breaking, and a nontrivial Berry phase were detected
\cite{1412.0824}.

The existence of surface Fermi arcs \cite{Savrasov,Haldane,Aji,Murakami} is
another fingerprint of Weyl semimetals, associated with the nontrivial topology.
Such arcs connect Weyl nodes of opposite chirality. The shape of the arcs
depends on the boundary conditions and can be engineered \cite{Hosur}.
The two Fermi arcs on opposite surfaces, together with the Fermi surface
of bulk states, form a closed Fermi surface. This implies, in particular, that
the chemical potentials for different chirality quasiparticles near distinct
Weyl points must be the same in a static system \cite{Haldane}.

Normally, one would not expect surface Fermi arcs in 3D Dirac
semimetals because the Berry flux vanishes for Dirac points with
vanishing topological charges. However, calculations of Refs.~\cite{Fang,WangWeng}
suggest that Dirac semimetals $\mathrm{A_3Bi}$ ($\mathrm{A}=\mathrm{Na}, \mathrm{K},
\mathrm{Rb}$) and $\mathrm{Cd_3As_2}$ possess nontrivial surface Fermi arcs.
This is an indication of a topologically nontrivial nature of such Dirac semimetals.
In fact, as we argue below, the situation is reminiscent of topological insulators, in
which there is a $\mathbb{Z}_2$ topological order associated with the time-reversal symmetry
\cite{Molenkamp,Hsieh,Kane,Hasan,Zhang}. This is further supported by the fact that
the breaking of time-reversal or inversion symmetry in Dirac semimetals causes splitting of the surface Fermi
arcs into a pair of open segments resembling the arcs in Weyl semimetals \cite{Fang}.
In this Rapid Communication, we explain the reason for the existence of nontrivial topological properties of
the $\mathrm{A_3Bi}$ compounds and shed light on their analytical structure in the
low-energy theory.

{\em Hamiltonian.}  Our starting point in the analysis will be the low-energy effective
Hamiltonian for electron excitations in $\mathrm{A_3Bi}$ ($\mathrm{A}=\mathrm{Na}, \mathrm{K},
\mathrm{Rb}$) derived in Ref.~\cite{Fang}. The explicit form of the Hamiltonian
is given by
\begin{eqnarray}
H(\mathbf{k}) &=& \epsilon_0(\mathbf{k}) + H_{4\times 4} ,
\label{WangWeng-Hamiltonian} \\
H_{4\times 4} &=& \left(
                                                    \begin{array}{cccc}
                                                      M(\mathbf{k}) & Ak_{+} & 0 & B^{*}(\mathbf{k}) \\
                                                      Ak_{-} & -M(\mathbf{k}) & B^{*}(\mathbf{k}) & 0 \\
                                                      0 & B(\mathbf{k}) & M(\mathbf{k}) & -Ak_{-} \\
                                                      B(\mathbf{k}) & 0 & -Ak_{+} & -M(\mathbf{k}) \\
                                                    \end{array}
                                                  \right).
\label{WangWeng-Hamiltonian4x4}
\end{eqnarray}
Note, that the Hamiltonian of the same form is also valid for structure I of 
$\mathrm{Cd_3As_2}$ (see Ref.~\cite{WangWeng}). The diagonal elements 
of the Hamiltonian are given in terms of two quadratic functions of
momenta,
$\epsilon_0(\mathbf{k}) = C_0 + C_1k_z^2+C_2(k_x^2+k_y^2)$ and
$M(\mathbf{k}) = M_0 - M_1 k_z^2-M_2(k_x^2+k_y^2)$.
The off-diagonal elements are determined by the functions $Ak_{\pm}$ and
$B(\mathbf{k}) = \alpha k_zk_{+}^2$, where $k_{\pm} = k_x\pm ik_y$.
The eigenvalues of Hamiltonian (\ref{WangWeng-Hamiltonian}) are
\begin{equation}
E(\mathbf{k})=\epsilon_0(\mathbf{k}) \pm \sqrt{M^2(\mathbf{k})+A^2k_{+}k_{-}+|B(\mathbf{k})|^2}.
\label{energy-dispersion}
\end{equation}
It is easy to check that the square root vanishes at the following two Dirac points:
$\mathbf{k}^{\pm}_0=\left(0, 0, \pm \sqrt{m}\right)$, where $m\equiv M_0/M_1$.
The function $B(\mathbf{k})$,
which can be interpreted as a momentum dependent mass term, also vanishes at
the Dirac points.

The general considerations of the current study will apply to the compounds
$\mathrm{A_3Bi}$ ($\mathrm{A}=\mathrm{Na}, \mathrm{K},\mathrm{Rb}$), but
our numerical results will be presented for $\mathrm{Na_3 Bi}$.
By fitting the energy spectrum of the effective Hamiltonian with the {\em ab initio}
calculations for $\mathrm{Na_3 Bi}$, the following numerical values of parameters
in the effective model were extracted \cite{Fang}:
$C_0 = -0.06382~\mbox{eV}$,
$C_1 = 8.7536~\mbox{eV\,\AA}^2$,
$C_2 = -8.4008~\mbox{eV\,\AA}^2$,
$M_0=-0.08686~\mbox{eV}$,
$M_1=-10.6424~\mbox{eV\,\AA}^2$,
$M_2=-10.3610~\mbox{eV\,\AA}^2$,
$A=2.4598~\mbox{eV\,\AA}$, and lattice constants are
$a=b=5.448~\mbox{\AA}$,
$c=9.655~\mbox{\AA}$.
Since no specific value for $\alpha$ was quoted in Ref.~\cite{Fang},
we will treat it as a free parameter below. For most of our analysis below, however,
the actual values of the model parameters are not very important. We will use
them only when presenting some numerical results.

In the simplest case of a vanishing mass function $B(\mathbf{k})$ (or, equivalently, for
$\alpha =0$), the Hamiltonian $H_{4\times 4}$ takes a block diagonal form:
$H_{4\times 4}(\alpha=0) \equiv H^{+}_{2\times 2}\oplus H^{-}_{2\times 2}$.
Its upper block is given by
\begin{eqnarray}
\label{Hamiltonians_{+}_New}
H^{+}_{2\times 2}=\left(
\begin{array}{cc}
        M(\mathbf{k}) & A(k_x+ik_y) \\
           A(k_x-ik_y) & -M(\mathbf{k})  \\
         \end{array}
       \right)
\end{eqnarray}
and has a very transparent physical meaning. It defines the simplest version of a Weyl
semimetal with two Weyl nodes located at $\mathbf{k}^{\pm}_0$. (The lower block
$H^{-}_{2\times 2}$ has a similar form, except that $k_x$ is replaced by $-k_x$.)
It is well known \cite{Murakami,Vishwanath} that such a Weyl semimetal has the
surface Fermi arc in the form of a straight line connecting the Weyl nodes of opposite
chirality at $\mathbf{k}^{+}_0$ and $\mathbf{k}^{-}_0$. Because of the sign difference,
$k_x \to -k_x$, the chiralities of the states near the Weyl nodes at $\mathbf{k}^{\pm}_0$
are opposite for the upper and lower block Hamiltonians. Thus, the complete
$4\times 4$ block diagonal Hamiltonian $H_{4\times 4}(\alpha=0)$ describes two
superimposed copies of a Weyl semimetal with two pairs of overlapping nodes.
The opposite chirality Weyl nodes coincide exactly in the momentum
space and, thus, effectively give rise to two Dirac points at $\mathbf{k}^{\pm}_0$.
At the same time, because the Weyl nodes come from different blocks, they cannot
annihilate and cannot form topologically trivial Dirac points. In fact, the corresponding
approximate model describes a $\mathbb{Z}_2$ Weyl semimetal. The nontrivial topological
properties, associated with the underlying $\mathbb{Z}_2$ Weyl semimetal structure, ensure that the
resulting Dirac semimetal possesses surface Fermi arcs.

It is easy to show that  the existence of the $\mathbb{Z}_2$ Weyl semimetal structure in this simplest
case is connected with the continuous symmetry $\mathrm{U}_{+}(1)\times \mathrm{U}_{-}(1)$
of the approximate Hamiltonian $H_{4\times 4}(\alpha=0)$. This symmetry describes
independent phase transformations of the spinors that correspond to the block Hamiltonians
$H^{+}_{2\times 2}$ and $H^{-}_{2\times 2}$, respectively.

{\em Symmetries.}  It is well known that, for $B(\mathbf{k})=\mbox{const}$, the
symmetry $\mathrm{U}_{+}(1)\times \mathrm{U}_{-}(1)$ is broken to its diagonal
subgroup $\mathrm{U}_{\rm em}(1)$
that describes the usual charge conservation. However, as we show below,
the low-energy Hamiltonian (\ref{WangWeng-Hamiltonian}) with the momentum
dependent mass function $B(\mathbf{k})=\alpha k_zk^2_{+}$ possesses a new
discrete symmetry that protects the $\mathbb{Z}_2$ Weyl semimetal structure.

Before discussing this symmetry, let us start by pointing out that the Hamiltonian
(\ref{WangWeng-Hamiltonian}) is invariant under the time-reversal and inversion
symmetries, i.e.,
\begin{eqnarray}
\Theta H_{-\mathbf{k}} \Theta^{-1} &=& H_{\mathbf{k}} \qquad \mbox{(time-reversal symmetry)},
\label{T-symmetry}
\\
PH_{\mathbf{-k}}P^{-1} &=& H_{\mathbf{k}} \qquad
\mbox{(inversion symmetry)},
\label{P-symmetry}
\end{eqnarray}
where $\Theta=TK$ ($K$ is a complex conjugation) and
\begin{equation}
T=\left(
       \begin{array}{cccc}
           0 & 0 & 1 & 0 \\
           0 & 0 & 0 & 1  \\
           -1 & 0 & 0 & 0 \\
           0 & -1 & 0 & 0
         \end{array}
       \right),
\quad
P=\left(
       \begin{array}{cccc}
           1 & 0 & 0 & 0 \\
           0 & -1 & 0 & 0  \\
           0 & 0 & 1 & 0 \\
           0 & 0 & 0 & -1
         \end{array}
       \right).
\label{T-P-matrices}
\end{equation}
Of course, these two symmetries are expected in Dirac semimetals such as 
$\mathrm{A_3 Bi}$, and they do play an important role in understanding
their physical properties. The less obvious is the following symmetry
defined by the transformation:
\begin{equation}
U H_{-k_z} U^{-1}= H_{k_z}  \quad \mbox{(ud parity)},
\label{flavor-symmetry}
\end{equation}
where matrix $U$ has the following block diagonal form: $U\equiv \mbox{diag}(I_2,-I_2)$
and $I_2$ is the $2\times 2$ unit matrix.
We call it the up-down (ud) parity because its eigenstates for $B(\mathbf{k})=0$ in view of
the block-diagonal structure of Hamiltonian (\ref{WangWeng-Hamiltonian4x4}) correspond to
bispinors with only two upper or lower nonzero components that describe a Weyl semimetal with a pair of
Weyl nodes. It should be noted that, for the Hamiltonian to be invariant under this symmetry,
it is crucial that the mass function $B(\mathbf{k})$ changes its sign when $k_z\to -k_z$
[while the functions $\epsilon_0(\mathbf{k})$ and $M(\mathbf{k})$ in the diagonal
elements do not change their signs]. Were the mass function momentum
independent, such a discrete symmetry would not exist.

The existence of the time-reversal (\ref{T-symmetry}) and ud parity
(\ref{flavor-symmetry}) symmetries has an important implication
that we will now explain. The argument relies on the fact that all
quasiparticle states in the low-energy model of a Dirac semimetal naturally
split into two separate groups, classified by the eigenvalues of the
operator $U_{\chi}=U\Pi_{k_z}$. (Here $\Pi_{k_z}$ is the operator that
changes the sign of the $z$ component of momentum, $k_z \to -k_z$.)
Taking into account that $U_\chi^2=1$, the eigenvalues
of $U_{\chi}$  are $\pm 1$. Furthermore, the corresponding eigenstates
are {\it not} invariant under time reversal. This follows from the fact that the
operators of time-reversal $\Theta$ and
ud parity $U_\chi$ transformations do not commute. This implies that
each group of quasiparticle states with a fixed eigenvalue of $U_\chi$ defines
a distinct copy of the Weyl semimetal, for which time reversal is broken.
Of course, the time-reversal symmetry is not broken in the
complete system including both $U_{\chi}$ sectors. In view of the $U_{\chi}$ symmetry,
we can classify the corresponding Dirac semimetal as a $\mathbb{Z}_2$ Weyl semimetal.
The situation resembles that of topological insulators \cite{Molenkamp,Hsieh,Kane,Hasan,Zhang},
which are time-reversal invariant due to the $\mathbb{Z}_2$ topological order parameter.

Each Weyl subsystem, described by quasiparticle states with a fixed eigenvalue of
$U_\chi$, has well defined Fermi arcs connecting the Weyl nodes at $\mathbf{k}^{\pm}_0$.
These arcs are topologically protected and cannot be removed by small perturbations of model
parameters.

In our discussion of Fermi arcs, it will be also useful to take into account that there
exists yet another discrete symmetry defined by the following transformation:
\begin{eqnarray}
\tilde{U} H_{-k_x} \tilde{U}^{-1} &=& H_{k_x},
\label{second-discrete-symmetry}
\end{eqnarray}
where
\begin{equation}
\tilde{U} = \left(
       \begin{array}{cccc}
           0 & 0 & 1 & 0 \\
           0 & 0 & 0 & 1  \\
           1 & 0 & 0 & 0 \\
           0 & 1 & 0 & 0
         \end{array}
       \right).
\label{U-prime-matrix}
\end{equation}
Of course, the product of the $U_{\chi}$ and $\tilde{U}\Pi_{k_x}$ transformations
$U_{\chi}\tilde{U}\Pi_{k_x}=T\Pi_{k_x}\Pi_{k_z}$ is also a symmetry of the low-energy
Hamiltonian (\ref{WangWeng-Hamiltonian}). Note that the symmetry $T\Pi_{k_x}\Pi_{k_z}$
is related to the time-reversal symmetry if we take into account that $K\Pi_{k_y}$
is also the symmetry of Hamiltonian (\ref{WangWeng-Hamiltonian}). Together the
operators $U_{\chi}$, $\tilde{U}\Pi_{k_x}$, and $T\Pi_{k_x}\Pi_{k_z}$ form a non-commutative
discrete group.

{\em Eigenstates of $U_{\chi}$.} Since Hamiltonian (\ref{WangWeng-Hamiltonian})
commutes with $U_{\chi}$, its eigenstates with eigenvalues $E(\mathbf{k})$ given
by Eq.(\ref{energy-dispersion}) can be chosen as eigenstates of $U_{\chi}$, too
(alternatively, we can choose the energy eigenstates to be eigenstates of the
$\tilde{U}\Pi_{k_x}$ or $T\Pi_{k_x}\Pi_{k_z}$ symmetries). These eigenstates
have the following form:
\begin{eqnarray}
\psi_{+}(\mathbf{k}) = N_{+} \left(
                 \begin{array}{c}
                   1 \\
                   \frac{E(\mathbf{k})-\epsilon_0(\mathbf{k})-M(\mathbf{k})}{Ak_{+}} \\
                  \frac{B(\mathbf{k})}{Ak_{+}} \\
                   0 \\
                 \end{array}
               \right), \\
\label{psi-plus}
\psi_{-}(\mathbf{k}) = N_{-}\left(
                 \begin{array}{c}
                  -\frac{B^{*}(\mathbf{k})}{Ak_{-}} \\
                   0 \\
                   1 \\
                   -\frac{E(\mathbf{k})-\epsilon_0(\mathbf{k})-M(\mathbf{k})}{Ak_{-}} \\
                 \end{array}
               \right).
\label{psi-minus}
\end{eqnarray}
Here $N_{\pm}$ are normalization constants and the subscript $\pm$ means the eigenvalue
of $U_{\chi}$. It is not difficult to check that $\tilde{U}\Pi_{k_x}$ transforms $\psi_+$ into
$\psi_-$ and vice versa. Notice that the bispinors $\psi_{\pm}$ in the case with a vanishing
mass function, $B(\mathbf{k})=0$, describe fermions of definite chirality in the vicinity of the
$\mathbf{k}^{\pm}_0$ points.

\begin{figure}[t]
\includegraphics[width=0.49\columnwidth]{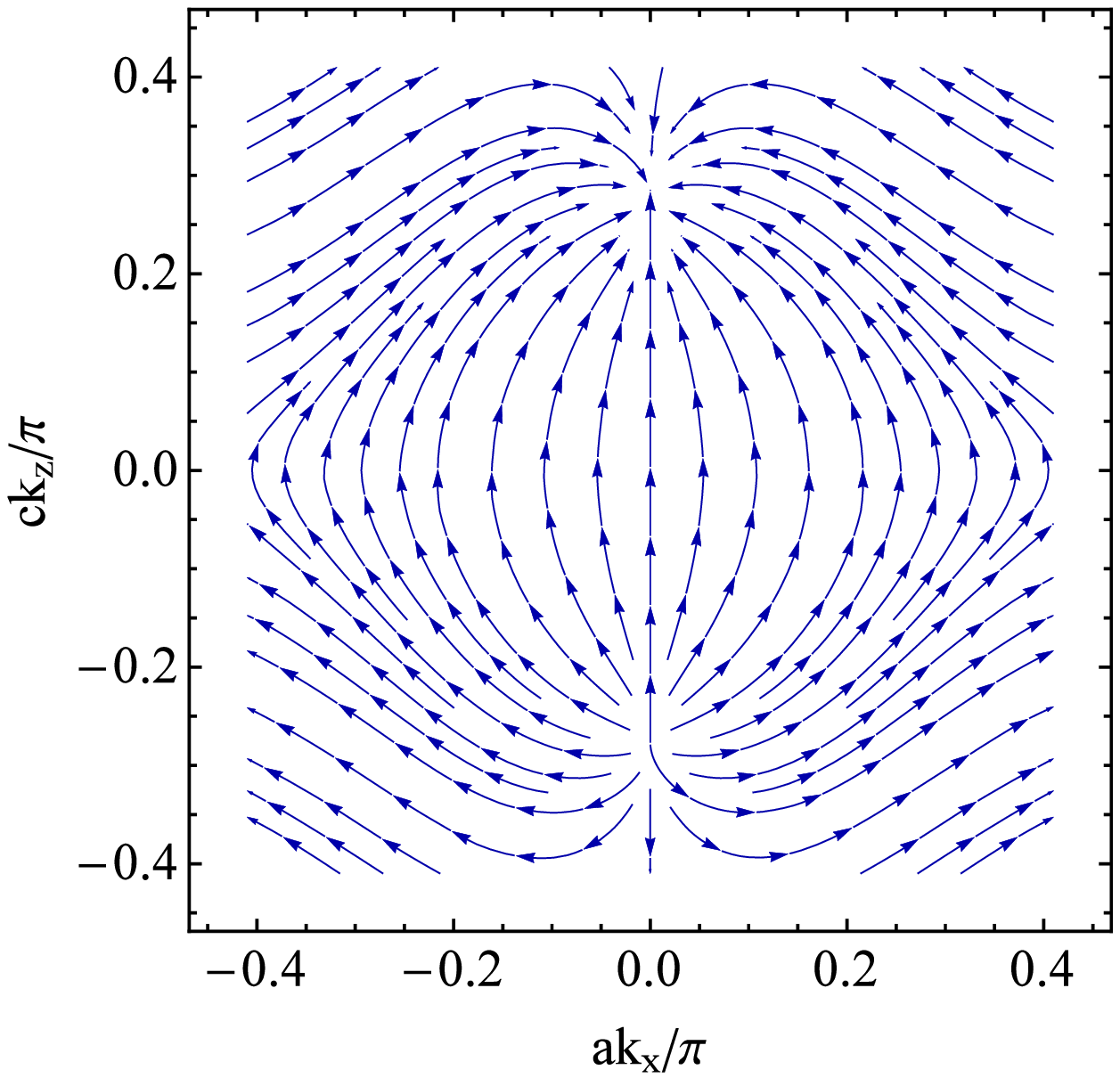}
\includegraphics[width=0.49\columnwidth]{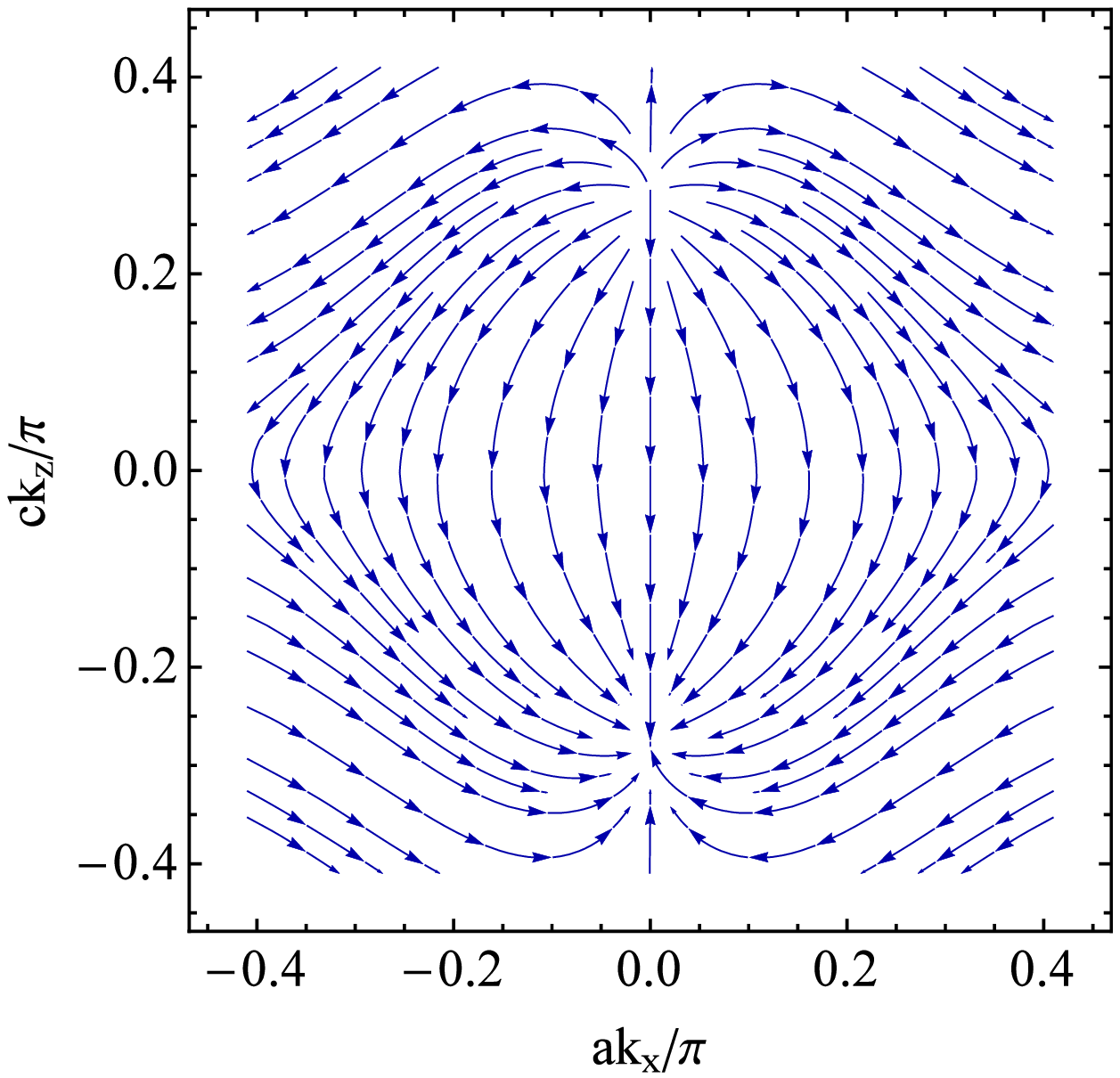}
\caption{(Color online) The projection of the Berry curvature $\mathbf{F}_{++}(\mathbf{k})$
(left panel) and $\mathbf{F}_{--}(\mathbf{k})$ (right panel) on the $k_y = 0$ plane.}
\label{fig:Berry-curvature}
\end{figure}

{\em Berry curvature.}
In order to explicitly reveal the $\mathbb{Z}_2$ Weyl semimetal  structure of $\mathrm{A_3 Bi}$
($\mathrm{A}=\mathrm{Na},\mathrm{K},\mathrm{Rb}$), we calculated the Berry connection
and the Berry curvature for each sector described by the $\psi_{\pm}(\mathbf{k})$ states.
Due to  the double degeneracy of the states with the same energy in the present case,
the Berry curvature is a matrix with non-Abelian gauge structure \cite{Wilczek}:
\begin{eqnarray}
\mathbf{A}_{mn}(\mathbf{k}) &\equiv& -\frac{i}{2}\left[\psi_{m}^{\dag}(\mathbf{k})
(\mathbf{\nabla}_{\mathbf{k}}\psi_{n}(\mathbf{k})) - (\mathbf{\nabla}_{\mathbf{k}}\psi_{m}^{\dag}(\mathbf{k}))
\psi_{n}(\mathbf{k})\right],  \nonumber\\
\label{Berry-connection}
\mathbf{F}_{mn}(\mathbf{k}) &\equiv& \mathbf{\nabla}_{\mathbf{k}}
\times \mathbf{A}_{mn}-i\mathbf{A}_{ml}\times \mathbf{A}_{ln}.
\label{Berry-curvature}
\end{eqnarray}
where $m,n,l=\pm$ and the summation over $l$ is performed in the last equation.
The four components of the Berry connection $\mathbf{A}_{mn}(\mathbf{k})$ define
a $\mathrm{U}(2)$ gauge field. The Berry curvature
components $\mathbf{F}_{++}(\mathbf{k})$ and $\mathbf{F}_{--}(\mathbf{k})$
are plotted in Fig.~\ref{fig:Berry-curvature}. The numerical results are shown
for $\alpha=50~\mbox{eV\,\AA}^3$ and the energy eigenvalue in
Eq.~(\ref{energy-dispersion}) with the positive sign in front of the square root.
(Up to the reflection $k_x\to -k_x$ and the change of direction of the vector fields,
the plots for the other sign of root look qualitatively the same.)

The results for the diagonal components of the curvature in Fig.~\ref{fig:Berry-curvature} 
show that each sector with a definite eigenvalue of $U_{\chi}$ contains a pair of Berry 
curvature monopoles with charges $\pm1$. Such a dipole structure in the momentum 
space is an unambiguous signature of a Weyl semimetal in each of the sectors.

We would like to emphasize that the presence of the mass function $B(\mathbf{k})$ does
not affect the property of the diagonal Berry curvature $\mathbf{F}_{++}(\mathbf{k})$
[or $\mathbf{F}_{--}(\mathbf{k})$] to have nonzero divergencies at the Weyl nodes.
Mathematically, the qualitative behavior of the curvature in the vicinity of the nodes
is preserved because $B(\mathbf{k})$ vanishes at $\mathbf{k}^{\pm}_0$. Away
from the Weyl nodes, on the other hand, the mass function does affect the
behavior of the Berry curvature. This is already seen in Fig.~\ref{fig:Berry-curvature},
where slight distortions of the dipole configurations become visible. It can be
checked that distortions become much stronger at larger values of parameter
$\alpha$. We found, however, that the opposite charge monopoles of the Berry
flux remain well resolved even for $\alpha$ as large as $250~\mbox{eV\,\AA}^3$.

It is interesting to point out that the off-diagonal components of the Berry
curvature $\mathbf{F}_{+-}(\mathbf{k})$ [as well as $\mathbf{F}_{-+}(\mathbf{k})$]
are nonzero only because of the nontrivial mass function $B(\mathbf{k})$. The
complete implications of this fact remain to be investigated. This task, however, is
beyond the scope of the present Rapid Communication.

{\em Surface Fermi arcs.} The nontrivial topological structure of the $\psi_+$ and $\psi_-$
sectors implies that the $\mbox{A}_3\mbox{Bi}$ compounds should have surface
Fermi arcs. Previously, the surface Fermi arcs in these 3D Dirac semimetals were
studied in Ref.~\cite{Fang} by using an iterative method to obtain the surface
Green's function of the semi-infinite system \cite{Yu}. The imaginary part of
the surface Green's function makes it possible to determine the local density
of states at the surface. In our study here, we employ the continuum low-energy
model and enforce appropriate boundary conditions for the quasiparticle spinors
at the surface of the semimetal. As we will argue, such a consideration
makes the physical properties of the surface Fermi arc states more transparent.

We assume that semimetal is situated at $y\ge0$ and is infinite in the $x$ and $z$
directions. The simplest implementation of the boundary condition for the semimetal
states on its surface follows from the replacement of $m$ with $-\tilde{m}$ and taking
the limit $\tilde{m}\to \infty$ on the vacuum side of the boundary \cite{Murakami}.
Taking into account that the Fermi arc states should be localized at the $y=0$ boundary,
we can look for the surface state solution in the following form:
\begin{equation}
\Psi(\mathbf{r})=\Psi_1 e^{-p_1y}+\Psi_2 e^{-p_2y},
\label{bottom-surface-semimetal-wave}
\end{equation}
where $\Psi_i$ can be chosen as the eigenstates of the $U_{\chi}$ symmetry
and $p_i$ are the positive (that are necessary for the normalization of the
wave function) roots of the characteristic equation
\begin{eqnarray}
&&\left[ C_2(k_x^2-p^2)+C_1k_z^2+C_0-E\right]^2+A^2 (p^2-k_x^2) \nonumber \\
&&-\left[M_0-M_1 k_z^2-M_2(k_x^2-p^2)\right]^2
-\alpha^2 k_z^2(p^2-k_x^2)^2=0.\nonumber \\
\label{char-eq-1}
\end{eqnarray}
The wave function on the vacuum side has a similar form, but with the replacement
$p_i\rightarrow-\tilde{p}_i$, where the definition of $\tilde{p}_i$ is similar to that of
$p_i$, but $m$ is replaced by $-\tilde{m}$. (In the calculation, we take the limit
$\tilde{m}\to \infty$, which prevents quasiparticles from escaping into vacuum.)

\begin{figure}[t]
\includegraphics[width=0.9\columnwidth]{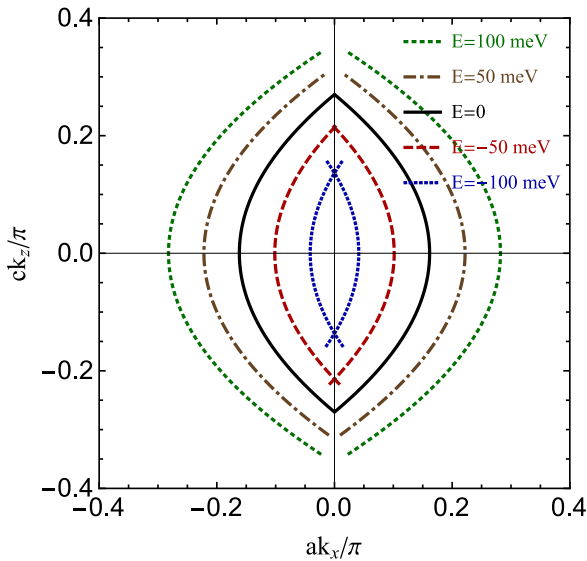}
\caption{(Color online) The surface Fermi arcs for $\alpha=1~\mbox{eV\,\AA}^3$
and energy $E=0,\pm 50,\pm 100$ $\,\mbox{meV}$.}
\label{fig:Fermi-Arc-5Es_alpha}
\end{figure}

Matching the wave functions across the boundary, we obtain the following equation:
\begin{equation}
\left(Q_1^{+} -Q_2^{+}\right)\left(Q_1^{-} -Q_2^{-}\right)-\left(T_1^{+} -T_2^{+}\right)\left(T_1^{-} -T_2^{-}\right)=0,
\label{Fermi-arcs-equation}
\end{equation}
where
\begin{eqnarray}
Q_{i}^{\pm} &=& -\frac{C_2(k_x^2-p_i^2)+C_1k_z^2+C_0-E \mp A k_x}{M_0-M_1k_z^2-M_2(k_x^2-p_i^2)-Ap_i},  \\
T_{i}^{\pm} &=& -\frac{\alpha k_z (p_i\pm k_x)^2} {M_0-M_1k_z^2-M_2(k_x^2-p_i^2)-Ap_i}.
\label{Q-T}
\end{eqnarray}
The numerical solutions for the surface Fermi arcs are shown in Fig.~\ref{fig:Fermi-Arc-5Es_alpha}
for several fixed values of the Fermi energy and $\alpha=1~\mbox{eV\,\AA}^3$. In the special case of $E=0$,
our results are in qualitative agreement with the results obtained in Ref.~\cite{Fang} by using a different method.

{\em Other materials.}
By combining the first-principles calculations and effective model analysis, it was
recently predicted \cite{Duan} that the ternary compounds $\mbox{Ba}Y\mbox{Bi}$
($Y=\mbox{Au},\mbox{Ag},\mbox{Cu})$ are Dirac semimetals. The low-energy effective
Hamiltonian of these compounds is similar to that of $\mbox{A}_3\mbox{Bi}$
($\mbox{A}=\mbox{Na},\mbox{K},\mbox{Rb}$), but with a different structure of the
mass terms
\begin{equation}
H_{\rm ternary} = \epsilon_0(\mathbf{k}) + H_{4\times 4}^\prime ,
\label{Hamiltonian-Duan}
\end{equation}
where
\begin{equation}
H_{4\times 4}^\prime = \left(
 \begin{array}{cccc}
 M(\mathbf{k}) & Ak_{+} & 0 & Bk_zk_{+}^2 \\
 Ak_{-} & -M(\mathbf{k}) & -Bk_zk_{+}^2 & 0 \\
 0 & -Bk_zk_{-}^2 & M(\mathbf{k}) & Ak_{-} \\
 Bk_zk_{-}^2 & 0 & Ak_{+} & -M(\mathbf{k}) \\
 \end{array}
\right).
\label{Hamiltonian-Duan4x4}
\end{equation}
Since Hamiltonian (\ref{Hamiltonian-Duan}) is invariant with respect to the $U_{\chi}$
symmetry transformation, our conclusions remain valid for these compounds. Thus,
these Dirac semimetals are $\mathbb{Z}_2$ Weyl semimetals too.

In conclusion, as we argued in this Letter, Dirac semimetals $\mbox{A}_3\mbox{Bi}$
($\mbox{A}=\mbox{Na},\mbox{K},\mbox{Rb}$) are, in fact, $\mathbb{Z}_2$ Weyl semimetals. The
conclusion is supported by the existence of the ud parity $U_{\chi}$ that allows us
to split all states into two sectors with each describing a Weyl semimetal. It is the
combination of both sectors that gives rise to a $\mathbb{Z}_2$ Weyl character of the corresponding
semimetals. Naturally, the time-reversal and inversion symmetries are preserved in such
a theory. The situation might be reminiscent of topological insulators, where the topological
order is protected by the time-reversal symmetry \cite{Molenkamp,Hsieh,Kane,Hasan,Zhang}.

The symmetry arguments used in the current study are rather powerful. They suggest
that the main conclusions should remain unchanged even in the presence of interaction effects,
provided the latter do not modify the low-energy spectrum in a qualitative way. A weak disorder
\cite{disorder} and a subcritical Coulomb interaction \cite{Coulomb} are the examples of such
effects that exist in realistic materials, but are not expected to change our main conclusions.

The fact that the compounds $\mathrm{A_3 Bi}$ ($\mathrm{A}=\mathrm{Na},\mathrm{K},
\mathrm{Rb}$) are $\mathbb{Z}_2$ Weyl semimetals has important implications. On the one hand, it
sheds light on the existence of surface Fermi arcs in such materials. This is particularly
important in view of the recent experimental confirmation of such states in $\mathrm{Na_3Bi}$
\cite{experimentNa3Bi}. Additionally, it predicts the same types of quantum oscillations as
in true Weyl semimetals \cite{Vishwanath}, with the period dependent on the thickness of
the semimetal slabs. Indeed, when the two Weyl sectors are protected from mixing by the
$\mathbb{Z}_2$ symmetry, their contributions will simply superimpose. Furthermore, when the
time reversal is broken (e.g., by magnetic impurities), we anticipate that the superposition
of two oscillations with nonequal periods will be observed.

The work of E.V.G. was supported partially by the Ukrainian State Foundation for Fundamental Research.
The work of V.A.M. was supported by the Natural Sciences and Engineering Research Council of Canada.
The work of I.A.S. was supported by the U.S. National Science Foundation under Grant No.~PHY-1404232
and in part by the Chinese Academy of Sciences Visiting Professorship for Senior International Scientists.

\end{document}